\documentclass[sigconf,nonacm]{acmart}

\usepackage{url}

\newcommand{\ARSName}{ARCADE}
\newcommand{\VideoLink}{https://youtu.be/9nH4c4Q-ooE}

\begin{document}

\title{\ARSName: An Augmented Reality Display Environment for Multimodal Interaction with Conversational Agents}

\author{Carolin Schindler}
\email{carolin.schindler@uni-ulm.de}
\orcid{0009-0007-5819-3035}
\affiliation{%
  \institution{Ulm University}
  \city{Ulm}
  \country{Germany}
}

\author{Daiki Mayumi}
\email{mayumi.daiki.mb9@is.naist.jp}
\orcid{0000-0002-6816-2132}
\affiliation{%
  \institution{Nara Institute of Science and Technology}
  \city{Ikoma}
  \country{Japan}
}

\author{Yuki Matsuda}
\email{yukimat@okayama-u.ac.jp}
\orcid{0000-0002-3135-4915}
\affiliation{%
  \institution{Okayama University}
  \city{Okayama}
  \country{Japan}
}

\author{Niklas Rach}
\email{niklas.rach@tensor-solutions.com}
\orcid{0000-0001-9737-8584}
\affiliation{%
  \institution{Tensor AI Solutions GmbH}
  \city{Ulm}
  \country{Germany}
}

\author{Keiichi Yasumoto}
\email{yasumoto@is.naist.jp}
\orcid{0000-0003-1579-3237}
\affiliation{%
  \institution{Nara Institute of Science and Technology}
  \city{Ikoma}
  \country{Japan}
}

\author{Wolfgang Minker}
\email{wolfgang.minker@uni-ulm.de}
\orcid{0000-0003-4531-0662}
\affiliation{%
  \institution{Ulm University}
  \city{Ulm}
  \country{Germany}
}

\renewcommand{\shortauthors}{Schindler et al.}

\begin{abstract}
  Making the interaction with embodied conversational agents accessible in a ubiquitous and natural manner is not only a question of the underlying software but also brings challenges in terms of the technical system that is used to display them.
  To this end, we present our spatial augmented reality system \ARSName{}, which can be utilized like a conventional monitor for displaying virtual agents as well as additional content.
  With its optical-see-through display, \ARSName{} creates the illusion of the agent being in the room similarly to a human.
  The applicability of our system is demonstrated in two different dialogue scenarios, which are included in the video accompanying this paper at
  \url{\VideoLink}.
\end{abstract}



\begin{teaserfigure}
  \centering
  \includegraphics[width=\textwidth]{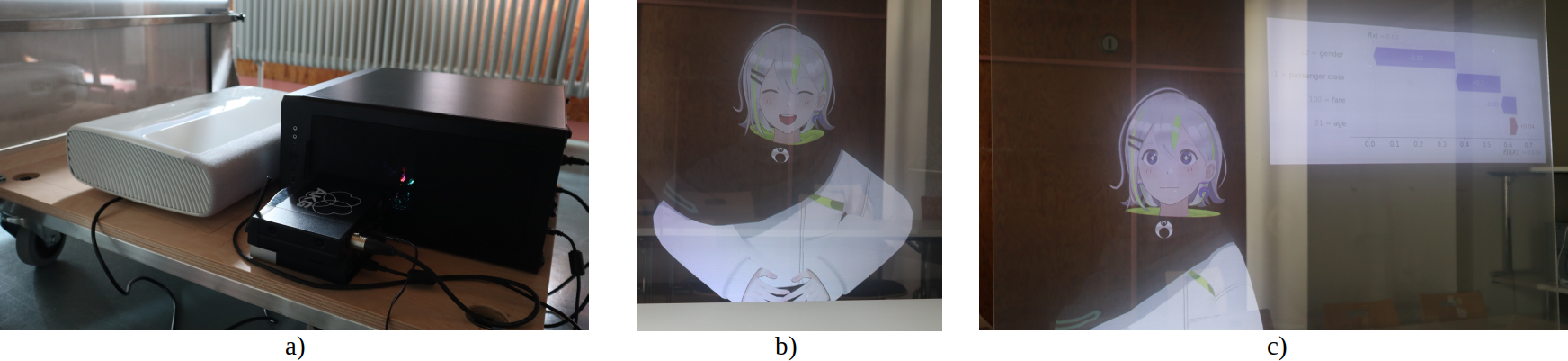}
  \caption{Impressions of ARCADE -- \textit{a)} Technical realisation, \textit{b)} Chit-chat dialogue, \textit{c)} XAI dialogue with additional visualisation}
  \Description{The figure is tripartite.
      Part A shows the hardware setup of ARCADE including the acrylic glass and projector.
      Part B shows an agent on ARCADE that seems to be sitting at a table. The agent is having a laughing expression.
      Part C shows an agent on ARCADE that has human-like size. Next to the agent, external content in the form of a plot is shown on ARCADE.
  }
  \label{fig:teaser}
\end{teaserfigure}

\maketitle

\section{Introduction}
Conversational systems are becoming increasingly ubiquitous in our everyday lives.
Embodying these systems by an agent~\cite{andre-pelachaud-2010-interacting}, allows for a rich multimodal interaction leading to a more human-like and hence intuitive interaction.
However, recent systems~(e.\,g.,~\cite{reinhardt-etal-2020-embedding, yamamoto-etal-2014-voice, aicher-etal-2023-influence}) do not integrate these virtual agents into the real environment in a truly natural manner.
Making use of augmented reality~(AR)~\cite{carmigniani-etal-2010-augmented}, the illusion of a virtual agent being present in the room similarly to a real person can be created.
Hence, it is worth investigating in the field of spatial AR, which integrates the AR experience directly into the environment and thus making it accessible to everyone passing by.

In this work, we present our in-room AR system,  \ARSName{}~(\underline{A}ug\-mented \underline{R}eality \underline{C}onversational \underline{A}gent \underline{D}isplay \underline{E}nvironment).
By allowing to project any 2D or 3D content onto a specially prepared transparent surface, \ARSName{} serves as an AR display for interacting more seamlessly and hence naturally with agents and respective underlying dialogue systems in the real world.
Our system can be used in the same way as a conventional monitor and thus is applicable to any conversational agent.

The remainder is organized as follows: In Section~\ref{sec:ARSystem}, we detail the design of \ARSName{} and demonstrate its applicability to embodied conversational agents, including showing additional information alongside the agent.
Section~\ref{sec:Conclusion} concludes the work with a brief summary and outlook.

\section{\ARSName{}}\label{sec:ARSystem}
\begin{figure}
  \centering
  \includegraphics[width=0.9\columnwidth]{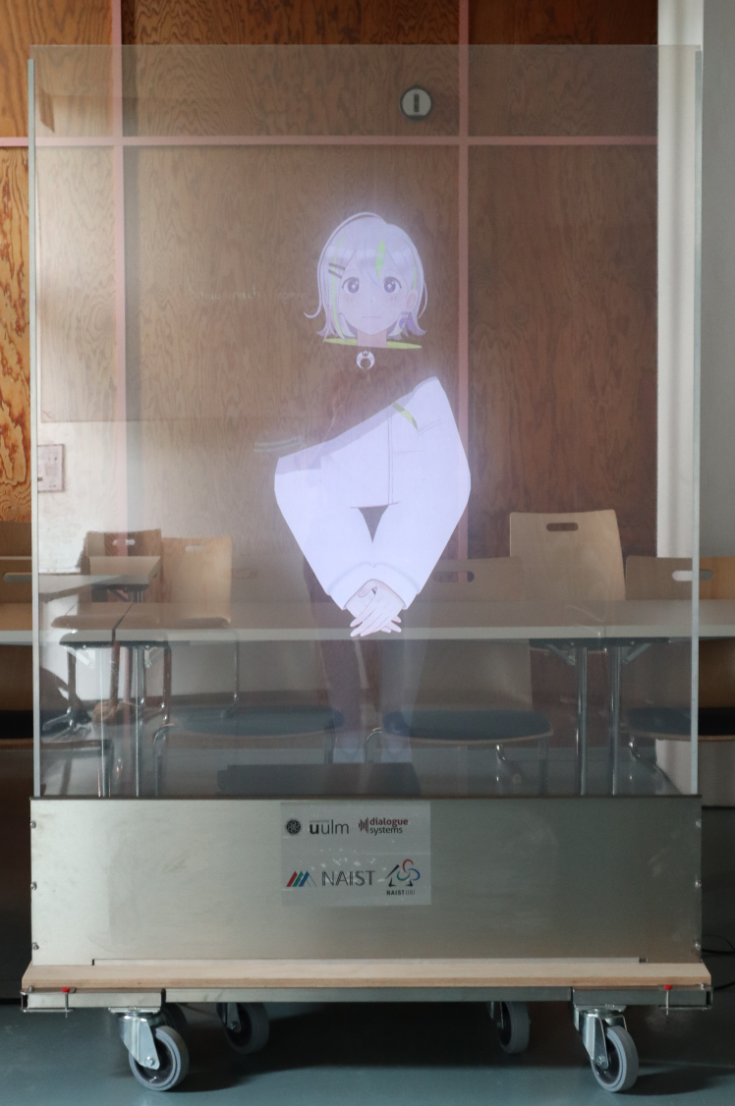}
  \caption{\ARSName{} from a user's perspective.}
  \Description{The figure shows the acrylic glass of \ARSName{} in full size with an agent projected onto it. At the bottom, the glass is mounted in some metal construction which is placed on a wooden platform on wheels.}
  \label{fig:overview}
\end{figure}

\ARSName{} is an augmented reality~(AR) system with a spatial optical-see-through display, allowing to interact with virtual agents of conversational systems in a natural and ubiquitous manner.
Figure~\ref{fig:overview} shows \ARSName{} in full size from a user's perspective.
In the following, we first summarize the requirements and features that we identified as being relevant for constructing~\ARSName{}.
Afterwards, we detail our technical realisation and exemplarily demonstrate the applicability to two different conversational agent systems.
A video of \ARSName{} including these exemplary demonstrations can be found at \url{\VideoLink}.

\subsection{Requirements and Features}
For our usage of \ARSName{} as an AR system for natural, human-like interaction with conversational agents, we define the following goals and corresponding requirements:
To create the illusion that the conversational agent is actually present in the room, every part of the display that is not occupied by the agent needs to resemble the real environment.
This also means that there should be no obvious border around the display.
For resembling human-human interaction, the agent needs to be displayed in the size of a real human.
Additionally, there should be the possibility to show external content, including texts, images, and videos, referenced during the conversation.
To account for ubiquitous usage, neither special environmental conditions nor special equipment is required by the user(s) to experience the illusion.
Furthermore, the effort of porting existing conversational agent systems to be displayed on \ARSName{} should be kept to a minimum.

\subsection{Technical Realisation}
The display is a transparent surface with a special foil for back-projection.
Where no content is projected, i.\,e., the color black is used, one can see the real environment through the transparent display.
To allow for real-sized virtual agents in sitting as well as standing scenarios, our surface is a $1.80$m high and $1.30$m wide acrylic glass.
The width moreover allows to display additional content on the surface next to the agent.
The acrylic glass is mounted on the bottom only, therewith preserving the boarder-less design of the system and hence the AR illusion.

We use a high luminosity, ultra-short distance projector to project onto the glass from behind.
The ultra-short distance property allows for a compact design of the system as shown in Figure~\ref{fig:teaser}a, while the high luminosity~($2800$~ANSI~Lumen) allows to use \ARSName{} by daylight as long as there is no source of light directly interacting with the beam of light of the projector.
For practical reasons, including storing and transporting the system in the laboratory, the system is on wheels and the part of the construction holding the acrylic glass can be separated from the remainder.

Any existing embodied conversational agent can be displayed through \ARSName{} by connecting the respective computer with the projector.
No specialized software or hardware link is required: Only the display output of the computer needs to be changed to the projector of the AR system.
This way, \ARSName{} is usable just like a conventional monitor and therefore can be integrated into existing research easily.

Since the illusion of the agent being in the room involves that it seems as if its utterances are spoken by it, the device for audio output should be placed close to the display.
In our case, we simply utilize the loudspeaker of the projector.
Depending on the scenario, additional sensors for recording and inferring the state of the user(s) can be utilized.
When the sensors need to be placed in the direction of the agent~(e.\,g.,~an RGB-D camera providing the system with vision), their positioning should be well thought out in order not to disrupt the illusion and the border-less design.

\subsection{Application to Exemplary Conversational Agents}
To demonstrate the applicability of \ARSName{} and to showcase the fulfillment of our requirements and features, we employ two different conversational agent systems.
Both systems are build upon the MMDAgent-EX~\cite{lee_2023_mmdagent} framework with ``Gene''\footnote{CG-CA Gene (c) 2023 by Nagoya Institute of Technology, Moonshot R\&D Goal 1 Avatar Symbiotic Society} as the embodiment of the agent.
However, every agent within any framework can be displayed on \ARSName{}.

\paragraph{Chit-Chat Dialogue with ChatGPT}
Following the implementation in REMDIS~\cite{chiba-etal-2024-remdis}, we employ ChatGPT\footnote{\url{https://openai.com/chatgpt/}} as the dialogue system manager behind the agent.
Based on the utterance by the user, ChatGPT is not only generating the agent's utterances but also selecting respective expressions.
The generated response texts are forwarded to the speech synthesis and the selected expressions are used to control the motion of the agent.
Thereby, the users can seamlessly engage in an every-day conversation with the agent in either a standing or, as in Figure~\ref{fig:teaser}b, a sitting scenario.

\paragraph{Explainable AI through Dialogue with Athena}
The field of explainable AI~(XAI)~\cite{das-rad-2020-opportunities}, which aims at providing transparency for decisions made by AI systems, is becoming more important the more complex and wide-spread AI systems are getting.
The dialogue system Athena~\cite{feustel-etal-2023-towards} offers access to these explanations in the form of a conversational interaction.
Its explanations are based on the XAI methods of counterfactuals~\cite{chou-etal-2022-counterfactuals} and Shapley values~\cite{merrick-taly-2020-explanation}.
Additionally, the explanations provided through the dialogue are supported by visual aids like a waterfall plot of the Shapley values.
Figure~\ref{fig:teaser}c shows an interaction with ``Gene'' embodying the Athena system and demonstrates having the plot as an additional source of context displayed next to the agent.

\section{Conclusion and Outlook}\label{sec:Conclusion}
We have presented \ARSName{}, an augmented reality system for natural and ubiquitous interaction with conversational agents.
With two conversational agents, one tailored towards multimodal chit-chat dialogue and the other towards explainable AI through dialogue, we demonstrated the applicability of \ARSName{} to different interaction scenarios and the fulfillment of our above defined requirements towards the system.
To improve the multimodal capabilities of the system towards symmetric multimodality~\cite{wahlster-2003-towards}, additional sensors for affective computing, among others an RGB-D camera, are going to be integrated into the system.
Moreover, it is of interest to evaluate the effect of our proposed AR system compared to existing ways of displaying embodied conversational agents in a user study.
Further work can be done in compressing the design of \ARSName{} to make it more portable; though, any transparent surface in an environment can be prepared as a display for \ARSName{}.

\begin{acks}
    We thank our colleagues Nicolas Wagner, Denis Dresvyanskiy, and Danila Mamontov at Ulm University, who provided valuable feedback during the planning and acquisition of \ARSName{}.
    Furthermore, we highly appreciate the support by our internal construction team at Ulm University, notably Thomas Kraft and Werner Birkle, in building up the system.
    Last but not least, we thank Sabrina Schindler for her essential contribution in creating the video accompanying this paper.
\end{acks}

\bibliographystyle{ACM-Reference-Format}
\bibliography{ARCADE_bibliography}


\begin{thebibliography}{12}


\ifx \showCODEN    \undefined \def \showCODEN     #1{\unskip}     \fi
\ifx \showDOI      \undefined \def \showDOI       #1{#1}\fi
\ifx \showISBNx    \undefined \def \showISBNx     #1{\unskip}     \fi
\ifx \showISBNxiii \undefined \def \showISBNxiii  #1{\unskip}     \fi
\ifx \showISSN     \undefined \def \showISSN      #1{\unskip}     \fi
\ifx \showLCCN     \undefined \def \showLCCN      #1{\unskip}     \fi
\ifx \shownote     \undefined \def \shownote      #1{#1}          \fi
\ifx \showarticletitle \undefined \def \showarticletitle #1{#1}   \fi
\ifx \showURL      \undefined \def \showURL       {\relax}        \fi
\providecommand\bibfield[2]{#2}
\providecommand\bibinfo[2]{#2}
\providecommand\natexlab[1]{#1}
\providecommand\showeprint[2][]{arXiv:#2}

\bibitem[Aicher et~al\mbox{.}(2023)]%
        {aicher-etal-2023-influence}
\bibfield{author}{\bibinfo{person}{Annalena Aicher}, \bibinfo{person}{Klaus Weber}, \bibinfo{person}{Elisabeth Andr{\'{e}}}, \bibinfo{person}{Wolfgang Minker}, {and} \bibinfo{person}{Stefan Ultes}.} \bibinfo{year}{2023}\natexlab{}.
\newblock \showarticletitle{The Influence of Avatar Interfaces on Argumentative Dialogues}. In \bibinfo{booktitle}{\emph{Proceedings of the 23rd {ACM} International Conference on Intelligent Virtual Agents, {IVA} 2023}}.
\newblock


\bibitem[Andr{\'e} and Pelachaud(2010)]%
        {andre-pelachaud-2010-interacting}
\bibfield{author}{\bibinfo{person}{Elisabeth Andr{\'e}} {and} \bibinfo{person}{Catherine Pelachaud}.} \bibinfo{year}{2010}\natexlab{}.
\newblock \showarticletitle{Interacting with Embodied Conversational Agents}. In \bibinfo{booktitle}{\emph{Speech Technology: Theory and Applications}}. \bibinfo{publisher}{Springer US}.
\newblock


\bibitem[Carmigniani et~al\mbox{.}(2010)]%
        {carmigniani-etal-2010-augmented}
\bibfield{author}{\bibinfo{person}{Julie Carmigniani}, \bibinfo{person}{Borko Furht}, \bibinfo{person}{Marco Anisetti}, \bibinfo{person}{Paolo Ceravolo}, \bibinfo{person}{Ernesto Damiani}, {and} \bibinfo{person}{Misa Ivkovic}.} \bibinfo{year}{2010}\natexlab{}.
\newblock \showarticletitle{Augmented reality technologies, systems and applications}.
\newblock \bibinfo{journal}{\emph{Multimedia Tools and Applications}} (\bibinfo{year}{2010}).
\newblock


\bibitem[Chiba et~al\mbox{.}(2024)]%
        {chiba-etal-2024-remdis}
\bibfield{author}{\bibinfo{person}{Yuya Chiba}, \bibinfo{person}{Koh Mitsuda}, \bibinfo{person}{Akinobu Lee}, {and} \bibinfo{person}{Ryuichiro Higashinaka}.} \bibinfo{year}{2024}\natexlab{}.
\newblock \showarticletitle{The Remdis toolkit: Building advanced real-time multimodal dialogue systems with incremental processing and large language models}. In \bibinfo{booktitle}{\emph{Proceedings of the 14th International Workshop on Spoken Dialogue Systems Technology (IWSDS '24)}}.
\newblock


\bibitem[Chou et~al\mbox{.}(2022)]%
        {chou-etal-2022-counterfactuals}
\bibfield{author}{\bibinfo{person}{Yu{-}Liang Chou}, \bibinfo{person}{Catarina Moreira}, \bibinfo{person}{Peter Bruza}, \bibinfo{person}{Chun Ouyang}, {and} \bibinfo{person}{Joaquim~A. Jorge}.} \bibinfo{year}{2022}\natexlab{}.
\newblock \showarticletitle{Counterfactuals and causability in explainable artificial intelligence: Theory, algorithms, and applications}.
\newblock \bibinfo{journal}{\emph{Inf. Fusion}} (\bibinfo{year}{2022}).
\newblock


\bibitem[Das and Rad(2020)]%
        {das-rad-2020-opportunities}
\bibfield{author}{\bibinfo{person}{Arun Das} {and} \bibinfo{person}{Paul Rad}.} \bibinfo{year}{2020}\natexlab{}.
\newblock \showarticletitle{Opportunities and Challenges in Explainable Artificial Intelligence {(XAI):} {A} Survey}.
\newblock \bibinfo{journal}{\emph{CoRR}} (\bibinfo{year}{2020}).
\newblock


\bibitem[Feustel et~al\mbox{.}(2023)]%
        {feustel-etal-2023-towards}
\bibfield{author}{\bibinfo{person}{Isabel Feustel}, \bibinfo{person}{Niklas Rach}, \bibinfo{person}{Wolfgang Minker}, {and} \bibinfo{person}{Stefan Ultes}.} \bibinfo{year}{2023}\natexlab{}.
\newblock \showarticletitle{Towards interactive explanations of machine learning methods through dialogue systems}. In \bibinfo{booktitle}{\emph{Proceedings of the 13th International Workshop on Spoken Dialogue Systems Technology (IWSDS '23)}}.
\newblock


\bibitem[Lee(2023)]%
        {lee_2023_mmdagent}
\bibfield{author}{\bibinfo{person}{Akinobu Lee}.} \bibinfo{year}{2023}\natexlab{}.
\newblock \bibinfo{booktitle}{\emph{{MMDAgent-EX}}}.
\newblock
\urldef\tempurl%
\url{https://doi.org/10.5281/zenodo.10427369}
\showDOI{\tempurl}


\bibitem[Merrick and Taly(2020)]%
        {merrick-taly-2020-explanation}
\bibfield{author}{\bibinfo{person}{Luke Merrick} {and} \bibinfo{person}{Ankur Taly}.} \bibinfo{year}{2020}\natexlab{}.
\newblock \showarticletitle{The Explanation Game: Explaining Machine Learning Models Using Shapley Values}. In \bibinfo{booktitle}{\emph{Machine Learning and Knowledge Extraction - 4th {IFIP} {TC} 5, {TC} 12, {WG} 8.4, {WG} 8.9, {WG} 12.9 International Cross-Domain Conference}}.
\newblock


\bibitem[Reinhardt et~al\mbox{.}(2020)]%
        {reinhardt-etal-2020-embedding}
\bibfield{author}{\bibinfo{person}{Jens Reinhardt}, \bibinfo{person}{Luca Hillen}, {and} \bibinfo{person}{Katrin Wolf}.} \bibinfo{year}{2020}\natexlab{}.
\newblock \showarticletitle{Embedding Conversational Agents into {AR:} Invisible or with a Realistic Human Body?}. In \bibinfo{booktitle}{\emph{{TEI} '20: Fourteenth International Conference on Tangible, Embedded, and Embodied Interaction}}.
\newblock


\bibitem[Wahlster(2003)]%
        {wahlster-2003-towards}
\bibfield{author}{\bibinfo{person}{Wolfgang Wahlster}.} \bibinfo{year}{2003}\natexlab{}.
\newblock \showarticletitle{Towards Symmetric Multimodality: Fusion and Fission of Speech, Gesture, and Facial Expression}. In \bibinfo{booktitle}{\emph{{KI} 2003: Advances in Artificial Intelligence, 26th Annual German Conference on AI}}.
\newblock


\bibitem[Yamamoto et~al\mbox{.}(2014)]%
        {yamamoto-etal-2014-voice}
\bibfield{author}{\bibinfo{person}{Daisuke Yamamoto}, \bibinfo{person}{Keiichiro Oura}, \bibinfo{person}{Ryota Nishimura}, \bibinfo{person}{Takahiro Uchiya}, \bibinfo{person}{Akinobu Lee}, \bibinfo{person}{Ichi Takumi}, {and} \bibinfo{person}{Keiichi Tokuda}.} \bibinfo{year}{2014}\natexlab{}.
\newblock \showarticletitle{Voice interaction system with 3D-CG virtual agent for stand-alone smartphones}. In \bibinfo{booktitle}{\emph{Proceedings of the second international conference on Human-agent interaction, {HAI} '14}}.
\newblock


\end{thebibliography}

\end{document}